# An interpretation for Aharonov-Bohm effect with classical electromagnetic theory

Gaobiao xiao, Shanghai Jiao Tong University, gaobiaoxiao@sjtu.edu.cn

**The magnetic Aharonov-Bohm effect shows that charged particles may be affected by the vector potential in regions without any electric or magnetic fields[1]. The Aharonov-Bohm effect was experimentally confirmed[2-3] and has been found in many situations[4-6]. A common explanation is based on quantum mechanics, which states that the wavefunctions associated with the charges will accumulate a phase shift due to the vector potential. However, consensus about its nature and interpretation has not been achieved[7-14]. We here propose a simple but reasonable interpretation based on the theory for electromagnetic radiation and couplings[15]. The energy associated with a pulse radiator is divided into a Coulomb-velocity energy and a radiative energy, together with a macroscopic Schott energy accounting for the energy exchange between them. All these energies are expressed with terms including the potentials, so are the mutual coupling energies. There exists a force acting on the moving charges even though the fields completely vanish. This force makes the charges pass through the magnetic solenoid in different velocity with different path length, causing a phase shift the same as that obtained with quantum mechanics. The theory is originally aimed for providing an interpretation for electromagnetic radiation and mutual coupling. It is derived directly from the Maxwell theory with no modification but only substitution and reorganization.**

The Aharonov-Bohm effect was first discussed in 1959[1]. As shown in Fig. 1, the magnetic field is confined within the magnetic solenoid. There are no magnetic fields or electric fields outside the solenoid. When charged particles pass the outside of the solenoid, there is a phase shift due to the nonzero static vector potential $\mathbf{A}(\mathbf{r})$. Consider two electron paths with the same start point and the same end point. One path goes in the upper side of the solenoid, the other in the lower side. The phase difference of the electrons traveling along the two paths have a phase difference dependent on the magnetic flux $\Phi$ in the solenoid,

$$\Delta\varphi = \frac{e}{\hbar}\oint \mathbf{A}\cdot\hat{l}dl = \frac{e}{\hbar}\Phi \quad (1)$$

Aharonov-Bohm effect is considered as an important quantum effect. It was experimentally confirmed by A. Tonomura[2-3], and has been found in various situations. Aharonov-Bohm effect implies that potentials may have a physical meaning instead of just a pure mathematical tool. However, this role of potentials has not been convincingly demonstrated[12]. The theory and interpretation for Aharonov-Bohm effect are still controversial[12,13].

This paper is not to discuss the difficulty involved in Aharonov-Bohm effect. Instead, we attempt to explain Aharonov-Bohm effect with macroscopic Maxwell theory as we believe that Aharonov-Bohm effect is a classical electromagnetic phenomenon that could be interpreted with classical electromagnetic theory. We have proposed a theory for electromagnetic radiation and mutual coupling[15], in which the energy of a pulse radiator is divided into three parts: a Coulomb-velocity energy that appears and disappears with its sources simultaneously, a radiative energy that leaves its sources after being emitted, and a reversible macroscopic Schott energy that is responsible for energy exchange. All energies are expressed with terms including potentials. The electromagnetic mutual couplings between two radiators can be handled in a similar way. With this formulation, the mutual coupling energies are not necessary to be zero when the force fields are all zeros. In the case of Aharonov-Bohm effect, the coupled electromagnetic energy between the electron and the solenoid can be evaluated with the vector potential instead of fields. The mutual energy varies with the motion of the electron, introducing a force on the electron to make it deflect, as shown in Fig.1. In the following we will show that this may possibly explain the Aharonov-Bohm effect in a simple but reasonable way.

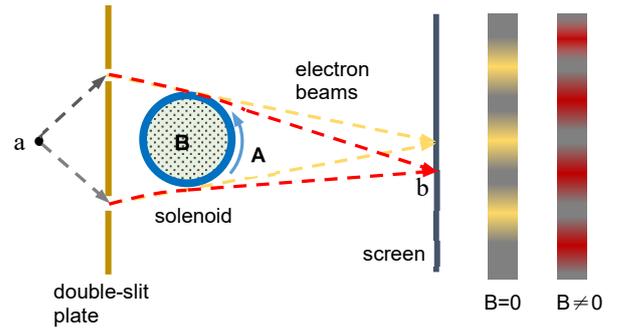

**Fig. 1 | Magnetic Aharonov-Bohm effect.** Coherent electrons from the two slits pass through a long solenoid from its upper and lower side, and form interference patterns on the screen. Without magnetic field in the solenoid, the interference pattern is like the yellow palette with bright strip at the center. With magnetic field, we show with the new interpretation that the electrons deflect due to mutual electromagnetic coupling, and form the interference patterns like the red palette. With the conventional interpretation based on quantum mechanics, the phase difference in the wave functions is directly related to the vector potential and causes the shift of the interference pattern, not necessarily requiring the deflection of the electrons.

## Energies of a pulse radiator

For a pulse radiator in vacuum in time period of 0≤*t*≤*T*, its energy can be divided into three parts[15,16] (Appendix A),

$$W_{tot}(t) = W_{\rho J}(t) + W_S(t) + W_{rad}(t) \quad (2)$$

Where we denote

$$W_{\rho J}(t) = \int_{V_s}\left[\frac{1}{2}\mathbf{J}(\mathbf{r},t)\cdot\mathbf{A}(\mathbf{r},t) + \frac{1}{2}\rho(\mathbf{r},t)\phi(\mathbf{r},t)\right]d\mathbf{r} \quad (3)$$

$$W_S(t) = \int_{V_\infty}\frac{1}{2}\frac{\partial}{\partial t}\left[\mathbf{D}(\mathbf{r},t)\cdot\mathbf{A}(\mathbf{r},t)\right]d\mathbf{r} \quad (4)$$

$$W_{rad}(t) = -\int_{V_\infty}\mathbf{D}(\mathbf{r},t)\cdot\frac{\partial \mathbf{A}(\mathbf{r},t)}{\partial t}d\mathbf{r} \quad (5)$$

In these equations, $\mathbf{A}(\mathbf{r},t)$ and $\phi(\mathbf{r},t)$ are respectively the vector and scalar potential at position $\mathbf{r}$ and time $t$ associated with current density $\mathbf{J}(\mathbf{r},t)$ and charge density $\rho(\mathbf{r},t)$ in domain $V_s$. $\mathbf{D}(\mathbf{r},t)$ is the electric flux density. The two potentials are subject to Lorentz Gauge and their zero points are at infinity.

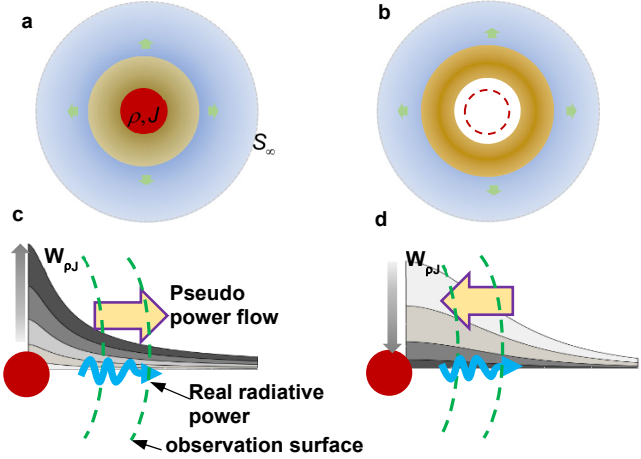

**Fig.2 | Schematic diagram of the energy distribution and transmission of a pulse radiator. a,** for 0≤$t$≤$T$, the Coulomb energy spreads over the whole space (blue area). The radiative energy spreads over the yellow region, the fluctuation of the Coulomb-velocity energy also occurs in this region. The radiative fields interact with sources in their way propagating outward through the source region (red area). There is a Schott energy responsible for energy exchange in this period. **b,** for $t \geq T+t_{max}/2$, the sources have disappeared and the total Coulomb-velocity energy becomes zero, but its fluctuation travels with radiative fields (yellow region). The total Schott energy is zero. The zero-field region (white area) expands with time. **cd,** the pseudo power flow due to the fluctuation of the Coulomb-velocity energy passes through the observation surface away from the source when the Coulomb-velocity energy increases, and toward source when the Coulomb-velocity energy decreases. The radiative power always crosses the observation surface away from the sources. The radiative fields travel to far region with a constant total radiative energy for $t \geq T+t_{max}/2$ until they encounter other sources.

The energies can be numerically evaluated with integrations over the source domains[15] from which their non zero duration in time domain can be determined. As illustrated in Fig.2, the macroscopic radiative source exists in the region within the dot circle. We assume that the source is static for $t$<0, becomes radiative for 0<$t$<$T$, and disappears for $t$>$T$. $W_{\rho J}(t)$ is the Coulomb-velocity energy associated with the Coulomb fields and the velocity fields. It appears and disappears with its sources simultaneously. For $t$<0, the static source has a Coulomb energy spreading in the whole space, and there is no velocity energy. For 0<$t$<$T$, the source is radiative and the Coulomb-velocity energy varies, with its fluctuation transmitting like waves with light velocity in free space. As shown in Appendix B, $W_S(t)$ is similar to the Schott energy in the charged particle theory, so it is defined as macroscopic Schott energy. It associates with radiative sources and is zero for static sources. Different from the Coulomb-velocity energy, the Schott energy does not disappear immediately with its radiative sources but continue to exist for a short while for $t \leq T+t_{max}/2$, where $t_{max}$ is the largest traveling time between two source points in $V_s$. $W_{rad}(t)$ is the radiative energy at time $t$. Its distribution region expands with the radiative wave to far region. However, the total radiative energy remains constant for $t \geq T+t_{max}/2$. There is a real radiative power flow passing through the observation surface always away from the source, together with a pseudo power flow due to the fluctuation of the Coulomb-velocity energy. The pseudo power flow may cross the observation surface forward if the Coulomb-velocity energy increases and backward if decreases. Note that all terms in the integrands in (3)-(5) consist of two quantities: the one in the left side is a source distribution or its fields, the one in the right side is the potential.

Consider a system of two sources $\rho_i(\mathbf{r}_i,t), \mathbf{J}_i(\mathbf{r}_i,t)$ in domain $V_{si}$, $i = 1,2$. The total energy of the system consists of self-energies ($i=j$) and mutual coupling energies ($i \neq j$),

$$W_{tot}(t) = \sum_{i=1}^{2}\sum_{j=1}^{2} W_{tot}^{ij}(t) = \sum_{i=1}^{2}\sum_{j=1}^{2} \left[ W_{\rho J}^{ij}(t) + W_S^{ij}(t) + W_{rad}^{ij}(t) \right] \quad (6)$$

where

$$W_{\rho J}^{ij}(t) = \int_{V_{si}} \left[ \frac{1}{2}\mathbf{J}_i(\mathbf{r}_i,t)\cdot\mathbf{A}_j(\mathbf{r}_i,t) + \frac{1}{2}\rho_i(\mathbf{r}_i,t)\phi_j(\mathbf{r}_i,t) \right] d\mathbf{r}_i \quad (7)$$

$$W_S^{ij}(t) = \int_{V_\infty} \frac{1}{2}\frac{\partial}{\partial t}\left[ \mathbf{D}_i(\mathbf{r},t)\cdot\mathbf{A}_j(\mathbf{r},t) \right] d\mathbf{r} \quad (8)$$

$$W_{rad}^{ij}(t) = -\int_{V_\infty} \mathbf{D}_i(\mathbf{r},t)\cdot\frac{\partial \mathbf{A}_j(\mathbf{r},t)}{\partial t} d\mathbf{r} \quad (9)$$

It can be interpreted that source-$j$ affects source-$i$ through its potentials. Note that if we consider the two radiators as a whole and treat them like a single radiator, (6)-(9) are completely in consistent with (2)-(5). Moreover, these formulae can be extended to systems consisting of multiple radiators.

An electromagnetic mutual coupling problem between two sources is shown in Fig.3**a**. Source-1 in $V_{s1}$ is static for $t \leq 0$ and $t > T$, and radiative for 0≤$t$≤$T$, while source-2 is static itself. Consider a small part of source (the yellow star in Fig.3a) in the source region $V_{s1}$. For 0≤$t$≤$T$, the radiative fields from the yellow star propagate outwards and interact with other sources that they have encountered in region $V_{s1}$ in their journey. Part of their radiative energy is transferred to those sources. Only after they have completely left the source region, the radiative fields from the yellow star can propagate with constant radiative energy, until they encounter source-2 and repeat the same interaction: change the Coulomb-velocity energy of source-2 and excite source-2 to emit radiative fields, as shown in Fig.3b. The macroscopic Schott energy is nonzero only in the energy exchanging stage. Although not illustrated in Fig.3, the variation of the Coulomb-velocity energy of the yellow star also affects the other sources in $V_{s1}$ and source-2 when the fluctuation reaches there. As has been discussed in previous section, the radiative energy of source-1 induces a real radiative power flow and the fluctuation the Coulomb-velocity energy of source-1 causes a pseudo power flow. They compose the total power flow that can be represented by the Poynting vector.

The analysis also reveals that the electromagnetic radiation and mutual coupling are the same issue and can be handled in the same way. Multiple sources can be collectively regarded as a single source with multiple nonconnected parts. The energies involved are the same as those in the radiation process, as expressed in (2)-(9).

The radiative fields of source-1 have no influence on source-2 when they have left $V_{s1}$ but have not yet reached $V_{s2}$. They just keep propagating in space and have no influence on sources in the space. However, when they have encountered any source in the space, all sources in the space may be affected soon or later through mutual couplings.

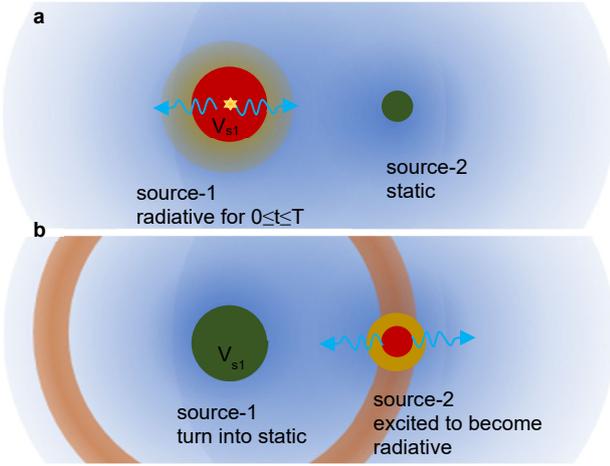

**Fig.3 | The radiative fields in electromagnetic radiation and mutual coupling.** Source-1 is radiative in 0≤t≤T and static for t≤0 and t≥T, while source-2 is static. Their Coulomb energy spreads over the whole space (blue area) for t≤0 and cause static mutual couplings. **a**, for 0≤t≤T, the radiative energy of source-1 has not yet reached source-2. The radiative fields from the yellow star interact with other sources which they have encountered in the way of traveling out of the source region $V_{s1}$. A nonzero macroscopic Schott energy exists in this period responsible for energy exchange. **b**, source-1 has turned back to be static. The radiative fields of source-1 have reached source-2 and interact with it. Source-2 could be excited to create a radiative energy and its Coulomb-velocity energy may change too. There is also a mutual macroscopic Schott energy in this period responsible for energy exchanging.

### Schott energy and radiative energy

Schott energy was introduced first by Schott[17] in 1912. The Schott energy of a moving charge $e$ in vacuum is (eq. (16) in Ref.18, eq. (4) in Ref.19)

$$E_S(t) = -\frac{1}{4\pi\varepsilon_0}\frac{2e^2}{3c^3}\mathbf{a}\cdot\mathbf{v} \tag{10}$$

where $\mathbf{v}(t)$ is its velocity, $\mathbf{a}(t)$ is its acceleration, $c$ and $\varepsilon_0$ are respectively the light velocity and permittivity in the vacuum. As shown in Appendix B, by applying the Lienard-Wiechert potentials[20,21] to the moving charge, the macroscopic Schott energy has the same form of (10). Making use of the relationship $\mathbf{E} = -\nabla\phi - \partial\mathbf{A}/\partial t$ and Lorentz Gauge $\nabla\cdot\mathbf{A} + c^{-2}\partial\phi/\partial t = 0$, $W_S(t)$ can be converted to an integral of the potentials,

$$W_S(t) = -\frac{\varepsilon_0}{4}\frac{d^2}{dt^2}\int_{V_\infty}\left[c^{-2}\phi^2(\mathbf{r},t) + \mathbf{A}(\mathbf{r},t)\cdot\mathbf{A}(\mathbf{r},t)\right]d\mathbf{r} \tag{11}$$

Choosing a spherical coordinate system and expanding the potentials in power series of $(\mathbf{n}\cdot\mathbf{v})$, it can be verified that $W_S(t) \approx E_s(t)$ for $v = |\mathbf{v}| \ll c$. Based on this relationship, and noting that both $W_S(t)$ and $E_s(t)$ are full time derivatives, we consider that $W_S(t)$ is corresponding to $E_s(t)$, and is termed as macroscopic Schott energy.

### Interpretation for Aharonov-Bohm effect

As shown in Fig.4, an infinitely long magnetic solenoid with radius $r_{sol}$ is put in free space with its axis coinciding with the z-axis. The magnetic field is $\mathbf{B} = \mu_0 J\hat{z}$ for $r < r_{sol}$, and $\mathbf{B} = 0$ for $r > r_{sol}$, where $J$ is the static current density in the solenoid. The vector potential is found to be $\mathbf{A}_{sol} = (A_0/r)\hat{\boldsymbol{\varphi}}$, with $A_0 = 0.5\mu_0 J r_{sol}^2$. An electron with charge $e$ and rest mass $m$ passes the solenoid with velocity $\mathbf{v}(t)$. As the vector potential and the current density in the magnetic solenoid is static, the electromagnetic energy of the solenoid remains unchanged in the whole process. Except the fields of the electron itself, there is no external fields in its path, hence no Lorentz force acting on the electron. The kinetic energy of the electron will be changed due to the electro-magnetic mutual coupling between the electron and the solenoid. In this case, the electromagnetic energy of the electron can be expressed by

$$W(\mathbf{r},t) = W_{el}(\mathbf{r},t) + W_{coup}(\mathbf{r},t) \tag{12}$$

where $W_{el}(\mathbf{r},t)$ is the total electromagnetic energy of the electron itself, and $W_{coup}(\mathbf{r},t)$ is the energy coupled from the solenoid to the electron. $\mathbf{r}$ is the position of the electron on its trajectory at time $t$. The coupling energy $W_{coup}(\mathbf{r},t)$ can be considered as an external work $W_{ext}$ done on the electron by an external force. The energy balance equation on the electron is then[18],

$$\frac{dW_{ext}}{dt} = \frac{d}{dt}\left[E_K + E_S + E_{rad}\right] \tag{13}$$

which is the equation (17) in Ref. 18. $E_K = 0.5m\mathbf{v}\cdot\mathbf{v}$ is the kinetic energy. $E_S$ is the Schott energy given in (10). $E_{rad}$ is the radiative energy. Its time derivative is given by the famous Larmor formula,

$$\frac{dE_{rad}}{dt} = \frac{1}{4\pi\varepsilon_0}\frac{2e^2}{3c^3}\mathbf{a}\cdot\mathbf{a} \tag{14}$$

The mass of the electron is approximately expressed by (equation (2-2) in Ref. 21)

$$m = f\frac{1}{4\pi\varepsilon_0}\frac{e^2}{r_o c^2} \tag{15}$$

where $f$ is a numerical constant of order 1 and $r_o$ is the radius of the electron. For the moving electron, $|\mathbf{a}|r_o/c \ll |\mathbf{v}|$, $|\dot{\mathbf{a}}|r_o/c \ll |\mathbf{a}|$. Therefore, we have

$$\begin{cases}\dfrac{2e^2}{3c^3}|\mathbf{a}\cdot\mathbf{a}| \ll \dfrac{2e^2}{3c^2 r_o}|\mathbf{a}\cdot\mathbf{v}| \\ \dfrac{2e^2}{3c^3}|\dot{\mathbf{a}}\cdot\mathbf{v}| \ll \dfrac{r_o}{c}\dfrac{2e^2}{3c^2}|\mathbf{a}\cdot\mathbf{v}| \ll \dfrac{2e^2}{3c^2 r_o}|\mathbf{a}\cdot\mathbf{v}|\end{cases} \tag{16}$$

Noting that $\partial E_K/\partial t = m\mathbf{a}\cdot\mathbf{v}$, we can immediately obtain that

$$\frac{dE_{rad}}{dt} \ll \frac{dE_K}{dt}, \quad \frac{dE_S}{dt} \ll \frac{dE_K}{dt} \tag{17}$$

Consequently, we have

$$\frac{dW_{ext}}{dt} \approx \frac{dE_K}{dt} \tag{18}$$

In Aharonov-Bohm situation, $W_{ext} = W_{coup}$. It is further checked in Appendix C that $W_{coup}(t) = 0.5\mathbf{v}(t)\cdot\mathbf{A}_{sol}(\mathbf{r})$, hence, the power balance becomes

$$\frac{d}{dt}\left[\frac{1}{2}m\mathbf{v}(t)\cdot\mathbf{v}(t)\right] \approx \frac{d}{dt}\left[\frac{1}{2}\mathbf{v}(t)\cdot\mathbf{A}_{sol}(\mathbf{r})\right] \tag{19}$$

Assume that the electron travels along its trajectory from $(\mathbf{r}_1, t_1)$ to $(\mathbf{r}, t)$. Integrating (19) over time $[t_1, t]$ yields

$$mv^2(t) - mv^2(0) = e\mathbf{v}(t)\cdot\mathbf{A}_{sol}(\mathbf{r}) - e\mathbf{v}(0)\cdot\mathbf{A}_{sol}(\mathbf{r}_1) \tag{20}$$

We consider the de Broglie wave for the electron in a diffraction situation. Substituting the momentum of the electron $p = mv = \hbar k$ into (20) yields

$$k(t)v(t) - k(0)v(0) = \frac{e}{\hbar}\mathbf{v}(t)\cdot\mathbf{A}(\mathbf{r}) - \frac{e}{\hbar}\mathbf{v}(0)\cdot\mathbf{A}(\mathbf{r}_1) \quad (21)$$

where $k$ is the wavenumber. Note that on the trajectory of the electron we have $dl = vdt$, and $\mathbf{v}(t)\cdot\mathbf{A}(\mathbf{r}) = \mp v(t)\hat{l}\cdot\mathbf{A}(\mathbf{r})$ for $l_\pm$, where $l_+$ denote the upper path and $l_-$ the lower path. $k(t)v(t)dt = kdl$ represents the phase shift of the electron over $dl$. Integrating (21) over time along the two trajectories and subtracting the two results gives the phase difference of the electron traveling through the two paths,

$$\Delta\varphi = \varphi_+ - \varphi_- = \frac{e}{\hbar}\oint_l \mathbf{A}\cdot\hat{l}dl = \frac{e}{\hbar}\Phi \quad (22)$$

which exactly agrees with (1).

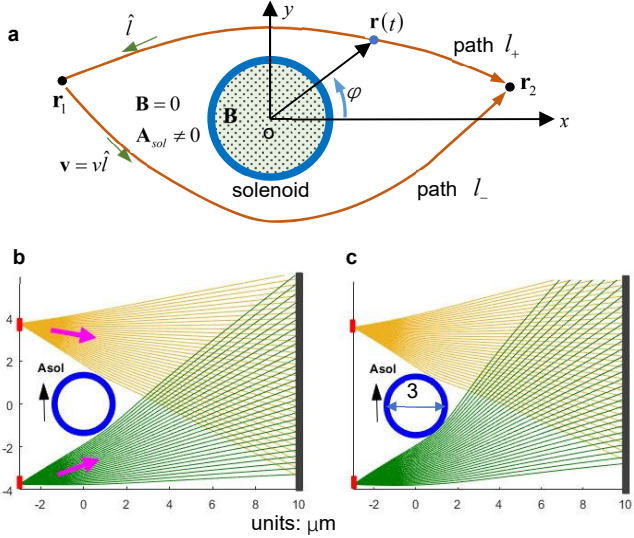

**Fig.4 | Deflection of electrons due to Aharonov-Bohm effect. a,** coordinate system. The axis of the infinite long magnetic solenoid is put on the z-axis, with an inner magnetic field in z-direction. An electron travels from point $\mathbf{r}_1$ to $\mathbf{r}_2$ along two paths: $l_+$ and $l_-$, with tangential unit $\hat{l}$. **b & c,** the trajectories of the electron from the two slits. $v(0)$=2cm/s for all electrons. Current density $J$ in the solenoid: 20mA/m for **b**, 40mA/m for **c**.

We have calculated the trajectories of the electrons for the system using finite difference method to solve (20) with an initial velocity of 2cm/s for all electrons. The results are shown in Fig.4b,c. As expected, larger deflection is observed for larger current in the solenoid.

**Discussions**

The theory proposed in Ref. 15 is originally aimed for interpreting the electromagnetic radiation and mutual coupling problems[22-26]. It has been shown here that a simple explanation for Aharonov-Bohm effect might be obtained based on the theory. Although there is no Lorentz force because of the absence of the fields, the analysis shows that the electrons indeed deflect due to the mutual electromagnetic coupling between the electrons and the solenoid, as has been predicted and confirmed by researchers[27-29]. We consider that this force is a kind of reactive force to the Lorentz force exerting on the current in the solenoid by the retarded magnetic field of the moving electron[7]. It can be justified by noting that for a moving electron in static vector potential, the energy coupled from the solenoid to the electron is not zero, i.e., $W_{react}^{coup}(t) + W_S^{coup}(t) \neq 0$, as shown in Appendix C. This coupling energy can be considered as the external work done on the electrons by the reactive force. If the solenoid is shielded by a super conductor[2], the mutual energy coupled from the solenoid to a moving electron is not affected since the vector potential of the solenoid is not affected and the energy in this case is evaluated very close to the electron, as shown in (45).

If the moving electrons are replaced by a current distribution $\mathbf{J}(\mathbf{r},t)$ outside the solenoid, we can show that the coupling energy can also be nonzero. Assume that the magnetic field and the electric flux density associated with $\mathbf{J}(\mathbf{r},t)$ are respectively $\mathbf{H}(\mathbf{r},t)$ and $\mathbf{D}(\mathbf{r},t)$. Recalling that $\nabla\times\mathbf{A}_{sol} = 0$ only at the region outside the solenoid, we can see that the energy coupled from the solenoid to the current source is not zero,

$$\begin{aligned} W_{react}^{coup}(t) &+ W_S^{coup}(t) \\ &= \int_{V_\infty}\frac{1}{2}\left[\mathbf{J}(\mathbf{r},t)+\frac{\partial}{\partial t}\mathbf{D}(\mathbf{r},t)\right]\cdot\mathbf{A}_{sol}(\mathbf{r})d\mathbf{r} \\ &= \int_{V_\infty}\frac{1}{2}\nabla\times\mathbf{H}(\mathbf{r},t)\cdot\mathbf{A}_{sol}(\mathbf{r})d\mathbf{r} \quad (23) \\ &= \int_{V_\infty}\frac{1}{2}\mathbf{H}(\mathbf{r},t)\cdot\nabla\times\mathbf{A}_{sol}(\mathbf{r})d\mathbf{r} \\ &= \int_{V_{sol}}\frac{1}{2}\mathbf{H}(\mathbf{r},t)\cdot\nabla\times\mathbf{A}_{sol}(\mathbf{r})d\mathbf{r} \neq 0 \end{aligned}$$

which may be interpreted that the magnetic fields of the source enter the solenoid and interact with the magnetic field inside the solenoid. If the solenoid is shielded by a super conductor, then the integration area is limited to the region outside the solenoid,

$$\begin{aligned} W_{react}^{coup}(t) &+ W_S^{coup}(t) \\ &= \int_{V_\infty-V_{sol}}\frac{1}{2}\left[\mathbf{J}(\mathbf{r},t)+\frac{\partial}{\partial t}\mathbf{D}(\mathbf{r},t)\right]\cdot\mathbf{A}_{sol}(\mathbf{r})d\mathbf{r} \\ &= \int_{V_\infty-V_{sol}}\frac{1}{2}\nabla\times\mathbf{H}(\mathbf{r},t)\cdot\mathbf{A}_{sol}(\mathbf{r})d\mathbf{r} \quad (24) \\ &= -\oint_{S_{sol}}\frac{1}{2}\left[\mathbf{H}(\mathbf{r},t)\times\mathbf{A}_{sol}(\mathbf{r})\right]\cdot\hat{\mathbf{n}}dS \\ &= \oint_{S_{sol}}\frac{1}{2}\left[\hat{\mathbf{n}}\times\mathbf{A}_{sol}(\mathbf{r})\right]\cdot\mathbf{H}(\mathbf{r},t)dS \end{aligned}$$

which could also be nonzero as $\mathbf{A}_{sol}(\mathbf{r})$ has nonzero tangential component on the surface.

The derivation reveals that the result of (22) is valid for magnetic solenoids with arbitrary shapes as long as the vector potential is time invariant. As a matter of fact, the phase difference exists for any pair of paths that form a closed loop circling the solenoid. As the traveling time of the electron in different paths depends not only on the path length but also on the velocity, it is not necessary to bring a time delay between two electrons traveling through two different paths. Therefore, the phase difference is not explicitly dependent on the time delay as observed by researchers[29].

The interpretation could be applied for explaining the Aharonov-Bohm effect in cases with magnetic whiskers or rings[2,31], only to note that an additional phase difference by the magnetic flux has to be added to the phase of the electrons without the magnetic flux. If the phase difference is π, then the bright-dark strips in the interference pattern will exchange. Since the mutual electro-magnetic coupling energies expressed with potentials are also valid for time varying situations, it may provide a possible explanation for Aharonov-Bohm effect with time varying vector potentials. For Aharonov-Bohm effect in materials, we may replace the materials with equivalence sources at first, then handle the coupling problems with the equivalence sources in free space, and finally interpret the Aharonov-Bohm effect involved.

It can be seen from (3) that the mutual coupling energy for an electron in a static scalar potential is $e\phi(\mathbf{r})$, so the main part of the force acting on the electron is $e\nabla\phi(\mathbf{r}) = e\mathbf{E}$, which is exactly the Coulomb force. This implies that electrons cannot be affected with the scalar potential alone in the absence of electric fields. Hence, the first version of electric Aharonov-Bohm effect[1] may not exist unless it is described in other formulations.

**Appendix A: The energies of a pulse radiator**

The total electromagnetic energy in vacuum is

$$W_{tot}(t) = \int_{V_\infty} \frac{1}{2}(\mathbf{E}\cdot\mathbf{D} + \mathbf{H}\cdot\mathbf{B})d\mathbf{r} \quad (25)$$

Substituting $\mathbf{E} = -\nabla\phi - \partial\mathbf{A}/\partial t$, and $\mathbf{B} = \nabla\times\mathbf{A}$ into (17) yields

$$W_{tot}(t) = \int_{V_\infty}\left(\frac{1}{2}\phi\rho + \frac{1}{2}\mathbf{A}\cdot\mathbf{J}\right)d\mathbf{r} + \int_{V_\infty}\frac{1}{2}\frac{\partial}{\partial t}(\mathbf{D}\cdot\mathbf{A})d\mathbf{r} \\ -\int_{V_\infty}\mathbf{D}\cdot\frac{\partial\mathbf{A}}{\partial t}d\mathbf{r} + \oint_{S_\infty}\frac{1}{2}(\mathbf{A}\times\mathbf{H} - \phi\mathbf{D})\cdot\hat{\mathbf{n}}dS \quad (26)$$

For a pulse radiator in period 0≤*t*≤*T*, the surface integral in (26) is zero because the fields never reach $S_\infty$. So we get (2)-(5).

The potentials and the electric flux density are defined in their usual forms as follow,

$$\phi(\mathbf{r},t) = \int_{V_s}\frac{\rho(\mathbf{r}_1, t - R_1/c)}{4\pi\varepsilon_0 R_1}d\mathbf{r}_1 \quad (27)$$

$$\mathbf{A}(\mathbf{r},t) = \mu_0\int_{V_s}\frac{\mathbf{J}(\mathbf{r}_1, t - R_1/c)}{4\pi R_1}d\mathbf{r}_1 \quad (28)$$

$$\mathbf{D}(\mathbf{r},t) = -\varepsilon_0\nabla\phi(\mathbf{r},t) - \varepsilon_0\frac{\partial}{\partial t}\mathbf{A}(\mathbf{r},t) \\ = -\int_{V_s}\int_{-\infty}^{\infty}\rho(\mathbf{r}_1, t_1)\nabla G(t - t_1 - R_1/c)dt_1 d\mathbf{r}_1 \\ -\frac{1}{c^2}\int_{V_s}\int_{-\infty}^{\infty}\mathbf{J}(\mathbf{r}_1, t_1)\dot{G}(t - t_1 - R_1/c)dt_1 d\mathbf{r}_1 \quad (29)$$

where $\mu_0$ is the permeability in vacuum. The time domain Green's function can be expressed using the Dirac delta function,

$$G_1(\mathbf{r},\mathbf{r}_1; t - R_1/c) = \frac{\delta(t - R_1/c)}{4\pi R_1} \quad (30)$$

The mutual coupling macroscopic Schott energy can then be written as

$$W_S^{ij}(t) = \int_{V_\infty}\frac{1}{2}\frac{\partial}{\partial t}(\mathbf{D}_i\cdot\mathbf{A}_j)d\mathbf{r} \\ = \int_{V_\infty}\left\{\mu_0\int_{V_{sj}}\begin{bmatrix}-\int_{V_{si}}\int_{-\infty}^{\infty}\rho_i(\mathbf{r}_i,t_1)\nabla G_i dt_1 d\mathbf{r}_i \\ -c^{-2}\int_{V_{si}}\int_{-\infty}^{\infty}\mathbf{J}(\mathbf{r}_i,t_1)\dot{G}_i dt_1 d\mathbf{r}_i\end{bmatrix}\cdot\int_{-\infty}^{\infty}\mathbf{J}(\mathbf{r}_j,t_2)G_j dt_2 d\mathbf{r}_j\right\}d\mathbf{r} \quad (31)$$

where $G_{i,j} = G(t - t_{i,j} - R_{i,j}/c)$ and $R_{i,j} = |\mathbf{r} - \mathbf{r}_{i,j}|$. The following

explicit expression[15,24] is useful to evaluate the integrations for energies,

$$I = \int_{V_\infty} G(\tau_1 - R_1/c)G(\tau_2 - R_2/c)d\mathbf{r} = \frac{c^2}{8\pi r_{21}} \quad (32)$$

where $r_{21} = |R_1 - R_2|$. The integration domain for $(t_1, t_2)$ is determined in deriving (32)[15,24]. With careful manipulation, the macroscopic Schott energy can be transformed to an integration over the source domain,

$$W_S^{ij}(t) = -\frac{1}{8\pi\varepsilon_0}\int_{V_{si}}\int_{V_{sj}}\frac{1}{r_{ij}} \times \int_{t-r_{ij}/c}^{t}\left[\begin{array}{c}\rho(\mathbf{r}_i,\tau)\dot\rho\left(\mathbf{r}_j,2t-\tau-\frac{r_{ij}}{c}\right)\\+c^{-2}\mathbf{J}\left(\mathbf{r}_i,2t-\tau-\frac{r_{ij}}{c}\right)\cdot\mathbf{J}(\mathbf{r}_j,\tau)\end{array}\right]d\tau d\mathbf{r}_j d\mathbf{r}_i \quad (33)$$

For a pulse source in $0 \leq t \leq T$, it can be checked that the integral becomes zero when $t \geq T + r_{ij,\max}/2c$ or $t \leq r_{ij,\min}/2c$.

The other mutual coupling energies can be derived in the same way. For example, we can check that

$$W_{\rho J}^{ij}(t) = \frac{1}{8\pi\varepsilon_0}\int_{V_{si}}\int_{V_{sj}}\frac{1}{r_{ij}}\left[\begin{array}{c}c^{-2}\mathbf{J}_i(\mathbf{r}_i,t)\cdot\mathbf{J}_j\left(\mathbf{r}_j,t-\frac{r_{ij}}{c}\right)\\+\rho_i(\mathbf{r}_i,t)\rho_j\left(\mathbf{r}_j,t-\frac{r_{ij}}{c}\right)\end{array}\right]d\mathbf{r}_j d\mathbf{r}_i \quad (34)$$

**Appendix B: Schott energy**

The Lienard-Wiechert potentials[20,21] for a moving charge are,

$$\phi(\mathbf{r},t) = \frac{1}{4\pi\varepsilon_0}\left[\frac{ce}{Rc-\mathbf{R}\cdot\mathbf{v}}\right]_{t'} = \frac{e}{4\pi\varepsilon_0}\left[\frac{1}{R(1-\mathbf{n}\cdot\boldsymbol{\beta})}\right]_{t'} \quad (35)$$

$$\mathbf{A}(\mathbf{r},t) = \frac{\mu_0}{4\pi}\left[\frac{ec\mathbf{v}}{Rc-\mathbf{R}\cdot\mathbf{v}}\right]_{t'} = \frac{e}{4\pi\varepsilon_0 c}\left[\frac{\boldsymbol{\beta}}{R(1-\mathbf{n}\cdot\boldsymbol{\beta})}\right]_{t'} \quad (36)$$

where $\mathbf{R} = \mathbf{r}-\mathbf{x}(t')$, $R = |\mathbf{r}-\mathbf{x}(t')|$, and $\mathbf{x}(t')$ is a point on the trajectory of the moving charge. $t' = t - R/c$ is the retarded time. Note that the quantities at the righthand side of (35) and (36) are evaluated at $t'$. $\mathbf{v}(t') = d\mathbf{x}(t')dt'$, $\boldsymbol{\beta} = \mathbf{v}/c$, and $\mathbf{n} = \mathbf{R}/R$. For $v = |\mathbf{v}| \ll c$, the potential terms can be expanded to the second order in $(\mathbf{n}\cdot\boldsymbol{\beta})$,

$$\phi^2(\mathbf{r},t) \approx \left(\frac{e}{4\pi\varepsilon_0}\right)^2\left[\frac{1}{R^2}\left(1+2\mathbf{n}\cdot\boldsymbol{\beta}+3(\mathbf{n}\cdot\boldsymbol{\beta})^2\right)\right]_{t'} \quad (37)$$

$$\mathbf{A}\cdot\mathbf{A} \approx \left(\frac{e}{4\pi\varepsilon_0 c}\right)^2\left[\frac{\boldsymbol{\beta}\cdot\boldsymbol{\beta}}{R^2}\right]_{t'} \quad (38)$$

The integrand of (11) can be divided into two parts,

$$I_{po}(\mathbf{r},t) = c^{-2}\phi^2(\mathbf{r},t) + \mathbf{A}(\mathbf{r},t)\cdot\mathbf{A}(\mathbf{r},t)$$
$$= \left(\frac{e}{4\pi\varepsilon_0 c}\right)^2\frac{1}{R^2}\left[(1+2\mathbf{n}\cdot\boldsymbol{\beta}+3(\mathbf{n}\cdot\boldsymbol{\beta})^2 + \boldsymbol{\beta}\cdot\boldsymbol{\beta})\right]$$
$$= \left(\frac{e}{4\pi\varepsilon_0 c}\right)^2\frac{1}{R^2} + \left(\frac{e}{4\pi\varepsilon_0 c}\right)^2\frac{1}{R^2}\left[2\mathbf{n}\cdot\boldsymbol{\beta}+3(\mathbf{n}\cdot\boldsymbol{\beta})^2 + \boldsymbol{\beta}\cdot\boldsymbol{\beta}\right]$$
$$\triangleq I_{po1} + I_{po2} \quad (39)$$

We are now to evaluate the Schott energy and the rate of radiation of the moving electron. We assume that the electron has nonzero velocity in a very short time period of $[t'-dt', t']$, and $t' = t - r_o/c$. Here $r_o$ is a small distance from the electron at $\mathbf{x}(t')$. It can be checked that, at time $t$, $I_{po2}$ is nonzero only within a thin shell $V_{shell}$, which can be approximately modelled with a nonuniform spherical shell with radius $r_o$ to $r_o + c(1-\mathbf{n}\cdot\boldsymbol{\beta})dt$, as shown in Fig. 5. In the spherical coordinate system with origin locating at $\mathbf{x}(t')$, $R = r$. We can write

$$d\int_{V_\infty}I_{po2}(\mathbf{r},t)d\mathbf{r} = \int_{V_{shell}}I_{po2}(\mathbf{r},t)d\mathbf{r}$$
$$\approx \left(\frac{e}{4\pi\varepsilon_0 c}\right)^2\int_0^{2\pi}\int_0^{\pi}\int_{r_o}^{r_o+c(1-\mathbf{n}\cdot\boldsymbol{\beta})dt'}\times$$
$$\left\{\frac{1}{r^2}\left[(2\mathbf{n}\cdot\boldsymbol{\beta}+3(\mathbf{n}\cdot\boldsymbol{\beta})^2 + \boldsymbol{\beta}\cdot\boldsymbol{\beta})\right]\right\}r^2\sin\theta dr d\theta d\varphi \quad (40)$$
$$\approx \left\{\left(\frac{e}{4\pi\varepsilon_0 c}\right)^2\int_0^{2\pi}\int_0^{\pi}\left[(2\mathbf{n}\cdot\boldsymbol{\beta}+(\mathbf{n}\cdot\boldsymbol{\beta})^2 + \boldsymbol{\beta}\cdot\boldsymbol{\beta})\right]\sin\theta d\theta d\varphi\right\}cdt'$$

A convenient choice is to put $\boldsymbol{\beta}$ on the z-axis, then we have $\mathbf{n}\cdot\boldsymbol{\beta} = |\boldsymbol{\beta}|\cos\theta$ and the integrand is symmetrical with z-axis. By checking that $\int_0^{\pi}\cos\theta\sin\theta d\theta = 0$, and $\int_0^{\pi}\cos^2\theta\sin\theta d\theta = 2/3$, we can obtain

$$\int_{V_{shell}}I_{po2}(\mathbf{r},t)d\mathbf{r} = \frac{1}{4\pi\varepsilon_0}\frac{4e^2}{3\varepsilon_0 c}\boldsymbol{\beta}(t')\cdot\boldsymbol{\beta}(t')dt' \quad (41)$$

which also holds true when $r_o = 0$ and $t' = t$, while we evaluate the integration at time $t$ with no retardation. Therefore, (41) can be rewritten as

$$\int_{V_{shell}}I_{po2}(\mathbf{r},t)d\mathbf{r} = \frac{1}{4\pi\varepsilon_0}\frac{4e^2}{3\varepsilon_0 c}\boldsymbol{\beta}(t)\cdot\boldsymbol{\beta}(t)dt = \frac{1}{4\pi\varepsilon_0}\frac{4e^2}{3\varepsilon_0 c^3}\mathbf{v}\cdot\mathbf{v}dt \quad (42)$$

$I_{po1}$ is not directly related to the velocity. Its integration should be evaluated over the whole space instead of the shell. It is a constant and vanishes after differentiation with respect to time $t$. Therefore, it can be deduced that

$$W_S(t) = -\frac{\varepsilon_0}{4}\frac{d}{dt}\left(\frac{d}{dt}\int_{V_\infty}I_{po2}(\mathbf{r},t)d\mathbf{r}\right) \approx -\frac{1}{4\pi\varepsilon_0}\frac{2e^2}{3c^3}\mathbf{a}(t)\cdot\mathbf{v}(t) \quad (43)$$

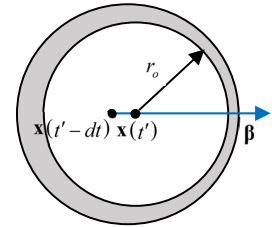

**Fig.5** | Integration region for evaluating $W_S(t)$.

**Appendix C: Estimation of energies in Aharonov-Bohm Effect**

The mutual energy coupled from the solenoid to the electron is

$$W_{coup}(t) = W_{\rho J}^{coup}(t) + W_S^{coup}(t) + W_{rad}^{coup}(t) \quad (44)$$

As the vector potential is static, and there is no scalar potential, making use of (7)-(9), we can show that the mutual coupled energy only contains two terms,

$$W_{coup}(t) = \frac{1}{2}e\mathbf{v}(t)\cdot\mathbf{A}_{sol}(\mathbf{r}) + \frac{1}{2}\frac{\partial}{\partial t}\int_{V_\infty}\mathbf{D}_{el}(\mathbf{r},t)\cdot\mathbf{A}_{Sol}(\mathbf{r})d\mathbf{r} \quad (45)$$

The second term in right hand side is the mutual macroscopic Schott energy. It can be estimated with the Lienard-Wiechert vector potential (36) and the nonrelativistic electric flux density of the electron,

$$\mathbf{D}(\mathbf{r},t) = \frac{e}{4\pi c^2} \left\{ \frac{c^2 (\mathbf{n}-\boldsymbol{\beta})}{R^2 (1-\mathbf{n}\cdot\boldsymbol{\beta})^3} + \frac{\mathbf{n}\times\left[(\mathbf{n}-\boldsymbol{\beta})\times\mathbf{a}\right]}{R(1-\mathbf{n}\cdot\boldsymbol{\beta})^3} \right\} \quad (46)$$

Since $\mathbf{A}_{sol}$ is static, $\partial\left[\mathbf{D}_{el}(\mathbf{r},t)\cdot\mathbf{A}_{Sol}\right]/\partial t = \mathbf{A}_{Sol}\cdot\partial\mathbf{D}_{el}(\mathbf{r},t)/\partial t$ is nonzero only on spheres with radius $r$ that satisfying $r = c(t-t')$. As the equation (18) should be balanced at the same time, we have to choose $r = r_o$. Consequently, the volume integral becomes a surface integral,

$$\int_{V_{shell}} \frac{\partial}{\partial t}\left[\mathbf{D}_{el}\cdot\mathbf{A}_{sol}\right]d\mathbf{r} = \oint_{S_o} \frac{\partial \mathbf{D}_{el}}{\partial t}\cdot\mathbf{A}_{sol}dS \quad (47)$$

in which the time derivative of the electric flux density can be obtained from (46). Keeping the terms including the first order of $\mathbf{a}$ or $\boldsymbol{\beta}$ gives

$$\frac{\partial}{\partial t}\mathbf{D}(\mathbf{r},t) = \frac{\partial t'}{\partial t}\frac{\partial}{\partial t'}\mathbf{D}(\mathbf{r},t)$$
$$\approx \frac{e}{4\pi c}\left\{\frac{c^2}{R}\left[3\mathbf{n}(\mathbf{n}\cdot\boldsymbol{\beta})-\boldsymbol{\beta}\right]-\mathbf{a}+3\mathbf{n}(\mathbf{n}\cdot\mathbf{a})\right\} \quad (48)$$

Taking $\mathbf{A}_{sol}$ as a constant vector on the small sphere, and making use of the relationships of $\int_0^\pi \cos\theta\sin\theta d\theta = 0$ and $\int_0^\pi \cos^2\theta\sin\theta d\theta = 2/3$ again, we can check that

$$W_S^{coup}(t) = \oint_{S_o} \mathbf{D}_{el}\cdot\mathbf{A}_{sol}dS \approx 0 \quad (49)$$